\documentclass[aps,prd,preprintnumbers,nofootinbib,twocolumn]{revtex4-2}
\usepackage{bm}
\usepackage{latexsym}
\usepackage{dcolumn}
\usepackage{amsmath,amsfonts,amssymb}
\usepackage{graphicx,float}
\usepackage{psfrag}
\usepackage{appendix}
\usepackage[hyperfootnotes=true]{hyperref}
\usepackage{mciteplus}
\usepackage{subcaption}
\usepackage{caption}
\usepackage{bbold}
\usepackage{slashed}
\usepackage{mathbbol}
\DeclareSymbolFontAlphabet{\amsmathbb}{AMSb}%
\usepackage{amsthm}
\usepackage{hyperref}
\usepackage{color}
\usepackage{mathrsfs}
\usepackage{setspace}
\usepackage{braket}

\DeclareMathAlphabet{\mathpzc}{OT1}{pzc}{m}{it}
\usepackage{calligra}
\DeclareMathAlphabet{\mathcalligra}{T1}{calligra}{m}{n}
\DeclareFontShape{T1}{calligra}{m}{n}{<->s*[2.2]callig15}{}

\def\be {\begin{equation}}
\def\ee {\end{equation}}
\def\bea {\begin{eqnarray}}
\def\eea {\end{eqnarray}}
\def\bc {\begin{center}}
\def\ec {\end{center}}
\def\bfg {\begin{figure}}
\def\efg {\end{figure}}
\def\bi {\begin{itemize}}
\def\ei {\end{itemize}}

\def\l  {\lambda}

\def\beq{\begin{equation}}
\def\eeq{\end{equation}}
\def\br{\begin{eqnarray}}
\def\er{\end{eqnarray}}
\newcommand{\eel}[1] {\label{#1}\end{equation}}

\emergencystretch=\maxdimen
\newcommand{\bdm}{\begin{displaymath}}
\newcommand{\edm}{\end{displaymath}}

\begin{document}


\title{At the Edge of Uncertainty: Decoding the Cosmological Constant value with Bose-Einstein Distribution}

\author{Ahmed~Farag~Ali$^{\triangle \nabla}$}
\email{aali29@essex.edu}

\author{Nader Inan$^{\Box   \oplus   \otimes }$}
\email{ninan@ucmerced.edu}

\affiliation{\small{$^\triangle$Essex County College, 303 University Ave, Newark, NJ 07102, United States.}}
\affiliation{\small{$^\nabla$Department of Physics, Faculty of Science, Benha University, Benha, 13518, Egypt.}}

\affiliation{$^\Box$ Clovis Community College, 10309 N. Willow, Fresno, CA 93730 USA}
\affiliation{$^\oplus$ University of California, Merced, School of Natural Sciences, P.O. Box 2039,
Merced, CA 95344, USA}
\affiliation{$^\otimes$ Department of Physics, California State University Fresno, Fresno, CA 93740-8031, USA}

\date{\today}

\begin{abstract}
\noindent
 We propose that the observed value of the cosmological constant may be explained by a fundamental uncertainty in the spacetime metric, which arises when combining the principle that mass and energy curve spacetime with the quantum uncertainty associated with particle localization. Since the position of a quantum particle cannot be sharply defined, the gravitational influence of such particles leads to intrinsic ambiguity in the formation of spacetime geometry. Recent experimental studies suggest that gravitational effects persist down to length scales of approximately $10^{-5}$ m, while quantum coherence and macroscopic quantum phenomena such as Bose-Einstein condensation and superfluidity also manifest at similar scales. Motivated by these findings, we identify a length scale of spacetime uncertainty, $L_Z \sim 2.2 \times 10^{-5}$ m, which corresponds to the geometric mean of the Planck length and the radius of the observable universe. We argue that this intermediate scale may act as an effective cutoff in vacuum energy calculations. Furthermore, we explore the interpretation of dark energy as a Bose-Einstein distribution with a characteristic reduced wavelength matching this uncertainty scale. This approach provides a potential bridge between cosmological and quantum regimes and offers a phenomenologically motivated perspective on the cosmological constant problem.

\end{abstract}

\maketitle

\section{Introduction}

The cosmological constant problem is widely acknowledged as one of the deepest challenges in modern theoretical physics. In his seminal review, Weinberg \cite{Weinberg:1988cp} highlighted how quantum field theory (QFT), when equipped with a Planck-scale cutoff, predicts a vacuum energy density that exceeds observational values by an enormous factor. Over the years, numerous approaches have been explored to reconcile this discrepancy. For example, early ideas by Zel'dovich \cite{Zeldovich:1968ehl} proposed that vacuum fluctuations, when properly accounted for with gravitational effects, might be naturally suppressed. Other researchers, including Carroll \cite{Carroll:1998zi} and Padmanabhan \cite{Padmanabhan:2002ji}, have advanced models such as quintessence and dynamical dark energy, attempting to render the cosmological constant an evolving quantity rather than a fixed parameter. In addition, there have been extensive efforts to revise the standard renormalization procedures in QFT. Parallel to these developments in high-energy physics, the study of macroscopic quantum phenomena in condensed matter physics has provided fresh perspectives. Bose-Einstein condensation (BEC) is a prime example, where a large number of bosons coalesce into a single quantum state, yielding observable effects such as long-range coherence and healing lengths \cite{Andrews1997}. More recently, this line of inquiry has been extended to cosmology; for instance, some studies have suggested that dark energy itself might be interpreted as a BEC \cite{Das:2014agf}. The intriguing possibility that the coherence lengths measured in such condensates could relate to cosmological scales has motivated our approach.

In this paper, we propose that the effective cutoff used in vacuum energy calculations should be derived from the uncertainty in the spacetime metric—a quantity that naturally emerges when combining quantum mechanics with general relativity. Notably, our derivation yields a cutoff length of approximately $2.2\times10^{-5}$~m, which is significant because it represents the geometric mean between the Planck length and the radius of the observable universe \cite{Freidel:2022ryr,Ng:1999hm}. This “middle scale” suggests a regime where neither purely quantum gravitational effects nor large-scale classical behavior fully dominate, but instead, both contribute crucially to the structure of spacetime.

Our work builds upon this rich body of literature by providing a model in which dark energy is treated as a massless Bose-Einstein condensate. By applying the Bose-Einstein distribution, we obtain a finite vacuum energy density with an effective temperature around 41 K and a corresponding reduced wavelength that mirrors our derived cutoff scale. This approach not only addresses the cosmological constant problem but also offers potential insights into how QFT may be modified at macroscopic quantum scales. The literature discussed above underscores the relevance of our model and motivates further exploration into the interplay between quantum mechanics and cosmology.

\section{Cosmological constant and Uncertainty}
\par\noindent
Among the many unresolved puzzles in theoretical physics, the cosmological constant problem stands out for its extreme degree of fine-tuning. In his influential analysis, Weinberg illustrated this discrepancy by adopting the Planck momentum as a natural cutoff for vacuum energy calculations—an approach that yields a value vastly larger than what is actually observed \cite{Weinberg:1988cp}. In contrast, the present work takes a critical stance on this common assumption. Rather than treating the Planck scale as an unquestioned boundary, we explore whether its role as a cutoff is physically justified, especially given that it merely marks the scale at which quantum gravitational effects are expected to emerge. It's natural to question if the cutoff should be of the Planck scale order, as Weinberg proposed in \cite{Weinberg:1988cp}. However, we diverge from this perspective for two key reasons: \emph{Firstly}, our approach belongs to a fundamental philosophy of physics: we rely on observations of nature to lead us to the theory, not vice versa. The observational value of the cosmological constant directs us to the cutoff on the vacuum energy density. To quote Max Planck, "Experiment is the only means of knowledge at our disposal. Everything else is poetry, imagination." Once we have precisely determined the cutoff suggested by natural observation, we then embark on developing the physical theory that can account for this value - a process that forms the basis of our paper. \emph{Secondly}, dark energy does not interact with the standard model of elementary particles, nor is it a component of this model.
Dark energy, required for cosmic acceleration, cannot be made up of photons or standard bosons, as they exhibit positive pressure. Dark energy necessitates a pressure-to-density ratio, known as the equation of state parameter, to be negative. Suggestions of potential dark energy candidates such as hypothetical quintessence \cite{Zlatev:1998tr,Carroll:1998zi,Wang:1998gt} further support this argument. Therefore, it's not physically justified to assume the cutoff for dark energy to be the same as the Planck scale, which is common for standard model particles. Moreover, the Planck scale does not represent a unique cutoff in the formalism of QFT (QFT) and is not even predicted by the theory. The concept of a cutoff represents a boundary beyond which a theory might not be valid, and its exact value is not unique \cite{schwartz2014quantum}. Different theories such as Quantum Electrodynamics (QED) and Quantum Chromodynamics (QCD) propose their specific cutoffs. For example, QED exhibits a Landau pole where QED ceases to be valid \cite{Landau:1954cxv}. Meanwhile, QCD introduces an ultraviolet cutoff where the non-perturbative gauge theory becomes relevant which may be explained through approaches to quantum gravity such as the AdS/CFT correspondence \cite{Maldacena:1997re,Gubser:1998bc,Witten:1998qj}. In fields like Lattice QFT, the cutoff, commonly interpreted as the inverse lattice spacing, doesn't maintain a universally consistent value \cite{montvay1994quantum}. Ideally, one would aim to minimize the lattice spacing to approximate the ideal version of QCD, but this would necessitate infinite computational resources. As such, a careful balance must be struck between achieving computational feasibility and maintaining accuracy when deciding the value for lattice spacing. The systematic errors originating from finite lattice spacing in lattice QCD calculations must be carefully considered. In this context, the cutoff method serves to quantify uncertainty or errors in the results, aligning perfectly with our paper's fundamental concept connecting cutoffs with uncertainty.  This work proposes a novel angle on the cosmological constant dilemma by treating the momentum cutoff not as a predetermined quantity, but as an unknown to be inferred from empirical cosmological data. Rather than assuming a fixed ultraviolet boundary, we interpret the cutoff as a threshold beyond which quantum field theory in a Lorentzian spacetime ceases to be reliable. In this framework, the cutoff acquires a physical interpretation: it reflects the intrinsic uncertainty in the spacetime geometry itself. In essence, we suggest that spacetime uncertainty defines the cutoff. To unpack this idea, we begin by revisiting how vacuum energy density is formally expressed.

\begin{eqnarray}
\rho_{\text{vac}} c^2 &=& \frac{1}{(2\pi \hbar)^3}\int^{P_Z}_0 (\tfrac{1}{2} \hbar \omega_p) ~ d^3p \\ &=& \frac{1}{16 \pi^3 \hbar^3} \int^{P_{Z}}_0 \sqrt{p^2 c^2+m^2 c^4} ~~ d^3p , \label{vacuum0}
\end{eqnarray}
In this formulation, \( P_Z \) denotes the momentum cutoff, and we adopt standard SI units throughout. It is crucial to note that Eq.~(\ref{vacuum0}) is derived under the assumption of a flat Lorentzian spacetime and does not incorporate any effects arising from spacetime curvature. Consequently, it is reasonable to expect that this equation remains valid only up to a certain scale—beyond which curvature effects become significant and the equation ceases to provide reliable results. The well-established principle that mass and energy influence the curvature of spacetime \cite{d1899introducing} asserts that energy distributions actively shape the geometric fabric of the universe. However, this classical insight encounters fundamental limitations when extended to quantum domains. In quantum mechanics, particles are not point-like with well-defined positions, but rather exist as probabilistic clouds governed by the uncertainty principle. As a result, any attempt to describe the spacetime curvature induced by such quantum entities must contend with an inherent ambiguity. This leads to a fundamental uncertainty in the associated spacetime geometry. From this viewpoint, it becomes natural to associate the cutoff momentum \( P_Z \) with the level of uncertainty in the metric. In other words, the momentum cutoff is not arbitrary—it reflects a physical limit arising from the quantum indeterminacy of spacetime itself. To explore this idea more precisely, let us consider the regime where \( mc \ll p \), or equivalently, where dark energy can be modeled by effectively massless particles. Under this assumption, Eq.~(\ref{vacuum0}) reduces to:
\begin{eqnarray}
{\rho_{\text{vac}} c^2= \frac{c}{16 \pi^3\hbar^3} \int^{P_{Z}}_0 p~(4\pi p^2 dp) = \frac{c}{16 \pi^2 \hbar^3} P_Z^4.} \label{vacuum}
\end{eqnarray}
De Broglie's duality gives
\begin{eqnarray}
{P_Z=\frac{\hbar}{L_Z},} \label{deBroglie}
\end{eqnarray}
where $L_Z$ is the cut-off length and can be understood as the reduced wavelength of de Broglie's wave \cite{Chang:2011jj}. Eq. (\ref{vacuum}) will be simplified to be
\begin{eqnarray}
\rho_{\text{vac}}c^2 = \frac{\hbar~c }{16 \pi^2 L_Z^4}
\label{vacuum1}
\end{eqnarray}
An essential question arises: What constitutes the uncertainty in spacetime? The answer can be found in several research works by Regge, Adler, Jack Ng and others \cite{Adler:2010wf,Regge:1958wr,Ng:1999hm,Ng:1994zk, Christiansen:2009bz, Mead:1964zz,Vilkovisky:1992pb,dewitt1964gravity,mead1966observable,garay1999quantum}. These studies have calculated the deviation from Lorentz geometry resulting from gravitational fields (spacetime curvature). In particular, it is shown in \cite{Adler:2010wf} that using $E=h v=h \left(\frac{c}{l}\right)$ and $\phi= \frac{Gm}{l}$ leads to
\begin{eqnarray}
{\Delta g=\frac{\phi}{c^2}=\frac{G\hbar}{c^3 \l^2}=\frac{{\ell_{Pl}}^2}{l^2}.} \label{curveuncertain}
\end{eqnarray}
If $\Delta g << 1$, then the spacetime is approximately Lorentzian. However, if $\Delta g=1,$ then there is significant spacetime uncertainty, also  known as "spacetime foam." 
As a result of Eq. (\ref{curveuncertain}), we see the existence of a degree of uncertainty in spacetime.

Now, we turn to Eq. (\ref{curveuncertain}) and consider $L_Z$ as the defining length where this uncertainty in Lorentz symmetry begins to emerge. This leads to
\begin{eqnarray}
{\Delta g= \frac{{\ell_{Pl}}^2}{L_Z^2}.} \label{uncertainmetric}
\end{eqnarray}
By substituting Eq. (\ref{uncertainmetric}) into Eq. (\ref{vacuum1}), we obtain
\begin{eqnarray}
{\rho_{\text{vac}}c^2= \frac{\hbar c}{16 \pi^2}\frac{\Delta g^2}{\ell_{Pl}^4} = 10^{74}\Delta g^2~ \frac{\text{(GeV)}^4}{(\hbar c)^3}.} \label{theory}
\end{eqnarray}
The \textit{observed} vacuum energy  found from the cosmological constant, 
$\Lambda \approx 10^{-52}$ m$^{-2}$, is:
\begin{eqnarray}
&\rho_{\text{observed}}& c^2=\frac{\Lambda c^{4}}{8\pi G}= 5.3 \times 10^{-10}~
\text{J/m}^{3}\qquad \label{rho_obs_with_CC}
\end{eqnarray}
or
\begin{eqnarray}
&\rho _{\text{observed}}& c^2= 2.57 \times 10^{-47}
\frac{\left( \text{GeV}\right) ^{4}}{\left( \hslash c\right) ^{3}}\label{rho_obs}
\end{eqnarray}
By comparing Eq. (\ref{theory}) with Eq. (\ref{rho_obs}), it becomes clear that the spacetime uncertainty can be found as
\begin{eqnarray}
{10^{74}~ \Delta g^2= 2.57 \times 10^{-47}} \implies
{\Delta g= 5\times 10^{-61}.}\label{metricuncertain}
\end{eqnarray}
In order to determine the value of $L_Z$, where the uncertainty in spacetime becomes relevant, we find that
\begin{eqnarray}
{L_Z^2=\frac{{\ell_\text{Pl}}^2}{\Delta g}= 5.14 \times 10^{-10} \implies
L_Z= 2.2\times 10^{-5}~m.}\label{L_z}\nonumber\\
\end{eqnarray}
The significance of the $2.2\times 10^{-5}~\text{m}$ value in Eq.~(\ref{L_z}) can be gleaned from studies by \cite{Freidel:2022ryr,Zeldovich:1968ehl,Tello:2023dqt}. These works, through various approaches such as the holographic principle and the entropy-area law \cite{tHooft:1993dmi,Susskind:1994vu,Bekenstein:1973ur}, suggest that this value equates to $L_Z\approx \sqrt{\ell_{\text{Pl}} \ell_u}$ \cite{Freidel:2022ryr}, where $\ell_u$ is the radius of the observable universe. This indicates that $2.2\times 10^{-5}~\text{m}$ is not a random length scale but is fundamentally tied to the physical parameters of our universe. Specifically, it could be the geometric mean of the smallest and largest known length scales in the universe.

This length scale, $L_Z \sim \sqrt{\ell_{\text{Pl}} \ell_u}$, is not merely coincidental but instead reflects a physically meaningful ``middle scale'' arising from the interplay between ultraviolet (UV) and infrared (IR) limits of the universe. Such geometric means are known to emerge in systems governed by two extreme scales, where effective phenomena manifest in the intermediate regime. In our case, $\ell_{\text{Pl}}$ sets the scale of quantum gravitational fluctuations, while $\ell_u$ defines the cosmic horizon—the IR bound for causal structure. The intermediate scale $L_Z$ has been previously proposed in holographic and information-theoretic contexts \cite{Ng:1999hm,Freidel:2022ryr}, where it marks the point at which spacetime ceases to behave classically but has not yet reached the Planckian regime. Furthermore, this scale closely matches the largest length at which macroscopic quantum coherence has been experimentally observed—such as in interference experiments of Bose-Einstein condensates \cite{Andrews1997} and the observation of quantum weirdness at the mesoscopic scale \cite{o2010quantum,kotler2021direct}. This coincidence supports the interpretation that $L_Z$ represents the \textit{maximum length scale for coherent quantum behavior} before gravitational backreaction or decoherence effects become non-negligible. Therefore, the emergence of $L_Z$ is not a numerical artifact but a profound signature of a mesoscopic regime where both quantum and gravitational effects coexist—a scale at which standard QFT cutoffs may naturally saturate without the need for ad hoc renormalization.

 The emergence of a new cutoff scale $L_Z \sim 2.2 \times 10^{-5}~\text{m}$, derived directly from the observed cosmological constant $\Lambda$, prompts a reevaluation of the standard assumptions in QFT. Traditionally, QFT assumes a flat Minkowski background with local Lorentz symmetry, and any cutoff scale—such as the Planck length—is treated as an ultraviolet (UV) regulator that is removed via renormalization. However, the existence of a physical, intermediate cutoff at $L_Z$, which reflects a transition region between quantum coherence and spacetime curvature effects, suggests that QFT must be modified in this mesoscopic regime. One natural direction is to promote the cutoff to a dynamical entity linked to spacetime uncertainty, thereby introducing a scale-dependent QFT formalism akin to curved-spacetime QFT but with built-in coherence loss at $L_Z$. This scale may mark the onset of gravitational decoherence and hence imply a breakdown of conventional renormalization group flows beyond $L_Z$. Furthermore, this framework would likely decouple dark energy from the Standard Model, suggesting that the vacuum structure of dark energy is governed by different statistical dynamics—such as a Bose-Einstein condensate phase—rather than by perturbative quantum fields. Such a perspective could lay the groundwork for a nonlocal or emergent field theory, where infrared physics (i.e., $\Lambda$) informs the ultraviolet behavior through holographic or entropic constraints \cite{tHooft:1993dmi,Ng:1999hm,Freidel:2022ryr}, bypassing the need for trans-Planckian extrapolations.

While our approach treats dark energy as a physically meaningful component responsible for the observed accelerated expansion of the universe, we acknowledge that the interpretation of cosmological observations within the $\Lambda$CDM model is not unique. Alternative cosmological models have been proposed that reproduce the observed data without invoking dark energy as a separate physical entity. For example, inhomogeneous cosmologies, modified gravity theories, and reinterpretations of cosmological redshift have been extensively discussed as potential alternatives \cite{Baryshev2012}. Furthermore, it is important to note that the existence of dark energy has not yet been confirmed by laboratory-based experiments. Our treatment therefore does not assume the ontological reality of dark energy, but rather aims to understand the observed value of the cosmological constant within a quantum framework, regardless of its ultimate interpretation.

A similar solution of cosmological constant problem was derived by one of the authors in \cite{Ali:2022ulp} based on universe radius and mass. Additional information regarding our proposal can be explored in detail in our recent essay \cite{Ali:2023gdc}. the principle ofmass/energy warping spacetime\cite{d1899introducing} proposes that the distribution of mass and energy shapes the fundamental structure of spacetime \cite{mach1960thescienceofmechanics, Einstein:1916vd}. The concept of quantum uncertainty  \cite{aHeisenberg:1927zz} hinders our ability to precisely determine the position of a quantum particle. This leads us to consider particles as mysterious clouds that conceal their exact location. These cloud-like particles don't just obscure the essential structure of spacetime where the particle moves, but they also suggest that our understanding of spacetime shaped by a quantum particle might not be complete because of the limitations in how we can observe reality. This idea of only seeing a piece of the reality was acknowledged by Einstein, Podolsky, and Rosen in \cite{Einstein:1935rr}. So, it seems that the cloudy nature of particles represents the uncertainty built into the measurement. The hydrogen atom, as shown in \cite{stodolna2013hydrogen}, is a clear example of this concept where both the electron and proton don't follow a clear trajectory.
\begin{figure}[htb]
\centering
\includegraphics[width=0.4\textwidth]{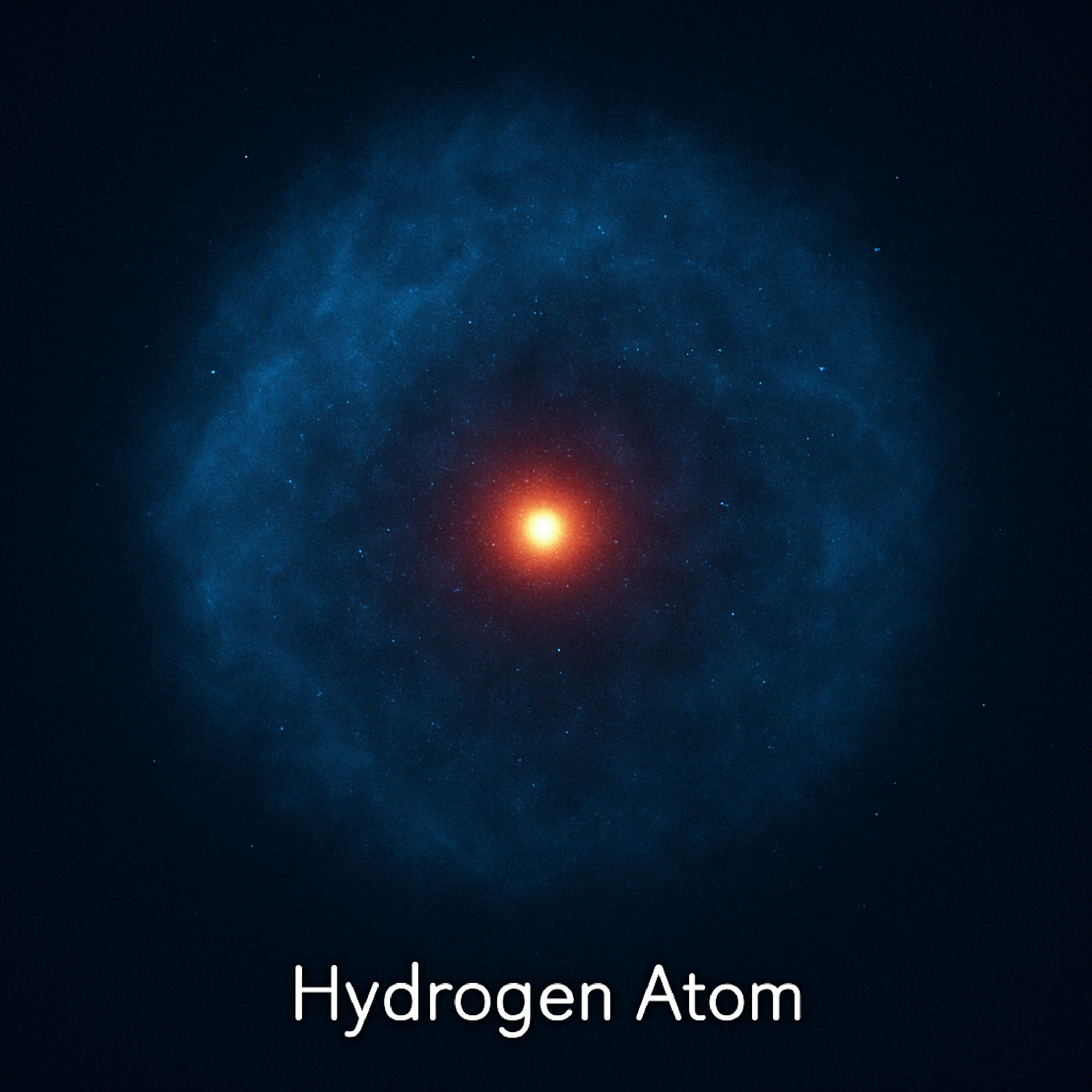}
\caption{Animated Image of Hydrogen atom cloud. Real image can be found in \cite{stodolna2013hydrogen}}
\label{cloud}
\end{figure}
The image of a hydrogen atom cloud (Figure \ref{cloud}) illustrates the constraints that quantum uncertainty imposes on our understanding of spacetime's true nature.
Hence, the length scale in Eq. (\ref{L_z})  suggests that quantum gravity begins to play a crucial role at a measurement of $2.2\times 10^{-5}~m$, which is the length scale at which quantum mechanics starts to be effective as well. 
Recent experiments have shown that gravity can still be effective at very small distances, even down to $5.2\times 10^{-5}~m$ \cite{Lee:2020zjt}. These results give strong support to our idea of applying the principle of mass/energy warping spacetime at the quantum level. This approach facilitates a deeper understanding of how the quantum particle clouds affect spacetime, consequently giving birth to the notion of spacetime uncertainty. However, QFT, which works in even smaller distances in a flat spacetime with assuming zero gravity, seems to conflict with these recent experiments \cite{Lee:2020zjt}. In QFT, we have to use a process called renormalization. Even though this method isn't mathematically rigorous, it is very useful in handling infinities. The idea of spacetime uncertainty could help to understand why the renormalization is necessary in QFT.

In a groundbreaking study, direct evidence of quantum weirdness was unveiled at a macroscopic scale of approximately $\approx  10^{-5}$~m by fabricating a quantum drum of that length \cite{o2010quantum,mercier2021quantum,kotler2021direct}.  Another experiment \cite{Andrews1997} demonstrated the macroscopic quantum phase of Bose-Einstein condensates. The observed interference pattern suggested a coherence length of approximately  $ \approx 10^{-5}$ m . Additionally, another theoretical study \cite{putterman1971macroscopic} seems to set a macroscopic limit on the applicability of the uncertainty principle by studying the flow of superfluid helium, It's suggested that long gravity wave propagation in helium II will be significantly restricted under a thickness of approximately $10^{-5}$ m. This is attributed to a macroscopic quantum uncertainty principle. A number of independent investigations have revealed that quantum uncertainty can manifest at unexpectedly large scales—the largest observed so far. Remarkably, this scale appears to be closely aligned with the characteristic length \( L_Z \), introduced in Eq.~(\ref{L_z}), which we derived from the uncertainty inherent in the spacetime metric and proposed as the natural cutoff for vacuum energy contributions. This convergence hints at a deeper connection: that the quantum uncertainty observed in particle behavior may ultimately stem from indeterminacy in the fabric of spacetime itself. Such a viewpoint implies that particles lack definite trajectories not merely due to quantum mechanical rules, but because the very geometry they inhabit is fundamentally uncertain. Within this framework, the approximate equivalence in magnitude between the macroscopic scale of observed quantum uncertainty and the proposed cutoff length \( L_Z \) is not coincidental—it may be pointing toward a resolution of the cosmological constant puzzle. That is, the vast discrepancy between theoretical predictions and observational values for the cosmological constant might originate from neglecting the role of metric uncertainty. In this light, quantum uncertainty, viewed as an emergent consequence of spacetime indeterminacy, could hold the key to understanding the true nature of vacuum energy.

Our approach also has implications for laboratory-based experiments that probe vacuum energy. Although the Casimir effect is often cited as experimental evidence for vacuum energy, it in fact measures differences in vacuum energy between boundary configurations, not its absolute value. Nevertheless, the introduction of a physical cutoff length scale such as $L_Z \sim 2.2 \times 10^{-5}$~m may provide a new lens for interpreting long-range quantum field fluctuations in such settings. For example, it could influence Casimir-type setups where cavity modes exceed this scale or suggest modifications in the spectral density of vacuum fluctuations at macroscopic distances. Moreover, our approach can be compared with several existing proposals aimed at resolving the cosmological constant problem. Anthropic arguments based on the string landscape \cite{Weinberg:1987dv}, dynamical scalar fields like quintessence \cite{Zlatev:1998tr}, unimodular gravity \cite{Unruh:1988in}, and renormalization group running of $\Lambda$ \cite{Shapiro:2000dz} each offer distinct frameworks. Unlike these models, our proposal introduces a physically motivated mesoscopic cutoff derived from experimental observations and geometrical arguments, without invoking extra dimensions, fine-tuning, or scalar degrees of freedom. This framework does not replace those efforts but complements them by highlighting the role of quantum uncertainty in shaping effective spacetime structure at cosmological scales.

\section{Dark Energy as a Massless Bose-Einstein Condensate}
\subsection{Renomarlization via the Bose-Einstein distribution function}
In the previous section, a  cutoff length scale was obtained by using the \textit{observed} vacuum energy density of the universe. The result was interpreted as a consequence of the length scale at which quantum spacetime uncertainty becomes relevant.
Intriguingly, recent experimental observations from Bose-Einstein condensates reveal that this cutoff length scale also mirrors the coherence length and healing length measured in these condensates \cite{Andrews1997,davis1995bose}. Such experimental findings suggest that these length scales potentially represent a form of the macroscopic uncertainty principle, consistent with the cutoff length scale deduced based on spacetime uncertainty.

Intriguingly, recent experimental observations from Bose-Einstein condensates reveal that this cutoff length scale also mirrors the coherence length and healing length measured in these condensates. Such experimental findings suggest that these length scales potentially represent a form of the macroscopic uncertainty principle, consistent with the cutoff length scale  deduced based on spacetime uncertainty. This similarity stimulates a fascinating proposition to perceive dark energy as a type of Bose-Einstein condensate.

Dark Energy, one of the most enigmatic entities in our understanding of the universe, has been previously modeled as a Bose-Einstein Condensate (BEC) owing to the unique macroscopic quantum behaviors demonstrated by BECs, including the macroscopic occupation of the ground state \cite{leggett2006quantum, pethick2008bose}. Unlike Fermi-Dirac statistics which do not permit such macroscopic occupation due to the Pauli Exclusion Principle, BECs comprise a large number of particles sharing the same quantum-mechanical state. This results in the manifestation of macroscopic quantum phenomena typically obscured on larger scales \cite{leggett2006quantum, pethick2008bose}. If dark energy were to behave as a BEC, we could anticipate a similar macroscopic occupation, suggesting a substantial portion of the universe's energy content to exist in a coherent, lowest-energy state. Furthermore, the superfluid nature of BECs, flowing without friction, could lend insight into the uniformity and isotropy of dark energy across the universe \cite{lifshitz2013statistical}. Critical temperature and density conditions characteristic of BEC formation might also provide hints about why dark energy possesses the energy density it does. The energy-momentum relationship of a particle in a Bose-Einstein condensate can be expressed as $E^2 = m^2 c^4 + p^2 c^2,$ where $E$ is the energy of the particle, $m$ the rest mass, and $p$ is the physical momentum. For massless particles, $m=0$, the relationship simplifies to $E =p c.$ In this framework, we understand the total vacuum energy density $\rho$ (i.e., the total vacuum energy per unit volume) by integrating the vacuum energy ($E= (1/2)~\hbar \omega$) over all momentum modes, with the function $f(p)$ representing the number of particles in each mode \cite{dodelson2020modern}. Then Eq. (\ref{vacuum0}) becomes
\begin{equation}
\rho c^2 = \frac{c}{16 \pi^3 \hbar^3} \int f(p) p \: d^3p\label{integral}\end{equation}
Assuming a thermal distribution for $f(p)$, we utilize the Bose-Einstein distribution\cite{pathria1996statistical}:
\begin{equation}
f(p) = \frac{g}{\exp((p c) / (k_B T)) - 1}
\end{equation}
Here, $g$ denotes the degeneracy factor, indicating the number of independent states that the particle can occupy \footnote{Although BEC’s are ordinarily composed of massive particles (such as atoms of rubidium, sodium, etc), the possibility of a BEC for massless particles is demonstrated by the fact that a BEC of photons has been achieved \cite{klaers2010bose}} . Solving the integral involves a change of variables and the use of the Bose-Einstein distribution function integral. By implementing the change of variable $x = p c / (k_B T)$, we obtain $p = k_B T x / c$ and $dp = k_B T dx / c$. In spherical momentum space, Eq. (\ref{integral}) becomes
\begin{equation}
\rho c^2 =\frac{g}{4 \pi^2 \hbar^3}  \left(\frac{k_B T}{c^{3/4}}\right)^4 \int_0^\infty \frac{x^3}{\exp(x) - 1} dx
\end{equation}
Evaluating the integral leads to
\begin{equation}
\rho c^2 = \frac{ g\pi^2}{60  (\hbar c) ^3}  \left(k_B T\right)^4
\end{equation}
The result can also be expressed as
\begin{eqnarray}
\rho c^2 = \alpha T^4 \label{energy}
\end{eqnarray}
where $\alpha = g \pi^2 k_B^4/(60 \hbar^3 c^3)$. This approach to the problem can be seen as an application of renormalization, eliminating the need for a cutoff. This is a concept that is quite similar to procedures in QFT (QFT), where we typically either assume a cutoff or apply renormalization procedures. Earlier research has proposed that the formation of dark matter could be attributed to the process of Bose-Einstein condensation \cite{Das:2014agf,Das:2022mgr,Boehmer:2007um}. Our current study extends this to explain the dark energy density or cosmological constant value. One may wonder why we consider modeling dark energy as a Bose-Einstein condensate and not as a system following Maxwell-Boltzmann or Fermi-Dirac statistics. The reason is deeply rooted in the statistical and quantum mechanical properties of the particles involved. In Maxwell-Boltzmann statistics, which is typically applicable to classical systems, particles are considered distinguishable. This characteristic implies that exchanging two particles would be considered a transformation to a different state of the system \cite{Huang1987}. In contrast, quantum statistics (such as Fermi-Dirac and Bose-Einstein) consider particles to be indistinguishable, meaning that swapping two particles does not result in a new state of the system \cite{mandl2010quantum}. This fundamental difference is crucial when discussing phenomena such as condensation. The principle of "indistinguishability," particularly relevant in the context of Bose-Einstein statistics, is involved in the formation of a Bose-Einstein condensation. When the temperature of a bosonic system approaches absolute zero, a significant fraction of particles begins populating the system's ground state \cite{pethick2008bose}. This arises from the property of bosons allowing multiple particles to occupy the same quantum state, culminating in the possibility of a macroscopic occupation of the ground state and hence a Bose-Einstein condensate \cite{pethick2008bose}. Although Maxwell-Boltzmann statistics do not impose restrictions on the number of particles occupying a given energy state, the assumption of particle distinguishability still persists \cite{Huang1987}. Hence, even with a high occupation number in lower energy states, it does not induce condensation analogous to that seen in a Bose-Einstein condensate. The distinguishability attribute of particles in the classical realm prevents them from displaying collective quantum mechanical behavior, as exhibited by bosons in a Bose-Einstein condensate. The classical particles maintain their individuality even at lower temperatures, and they continue to occupy a range of energy states following the Maxwell-Boltzmann distribution \cite{Huang1987}.

\subsection{Validating the model via consistency with General Relativity}

When studying the association between the cosmological constant and the energy density of a Bose-Einstein condensate (BEC), we reconcile two distinct equations for $\rho$. The first equation, Eq. (\ref{rho_obs_with_CC}), is derived from the Einstein field equations of General Relativity. The second equation, Eq. (\ref{energy}), is derived from relativistic QFT and modeling dark energy as a Bose-Einstein condensate. By equating these two results, we are then able to express $T$ in terms of the cosmological constant $\Lambda$, as follows:
\begin{eqnarray}
T = \left(\frac{\Lambda c^4}{8\pi G \alpha}\right)^{\frac{1}{4}}.
\end{eqnarray}
Applying known constants: $G = 6.67 \times 10^{-11}~ m^3 kg^{-1} s^{-2}$, $c = 299792458$~ m/s, $\Lambda= 1.1056 \times 10^{-52}~ m^{-2}$, $\hbar= 1.054 \times 10^{-34} ~ m^2 kg / s$, $k_B= 1.38 \times 10^{-23}~J/K$, and $g = 1$ (a single bosonic degree of freedom), allows us to deduce
\begin{equation}
T= 41~ K \label{temperature}
\end{equation}
Note that this temperature value is below the critical temperature necessary  to form a Bose-Einstein condensate from atoms (100 K). We must now ensure the compatibility of this temperature value with the cutoff derived in Eq. (\ref{L_z}). This particular temperature value should reproduce the length cutoff that we acquired in the preceding section. If it does not, then our hypothesis that considers Dark energy as a Bose-Einstein condensate would be flawed. Thus, we will calculate the corresponding length scale according to the standard relationship between energy and wavelength. 
\begin{eqnarray}
    k_B T= \frac{1}{2} \hbar \omega= \frac{1}{2} \frac{\hbar c}{ \lambdabar} \label{Tlambda}
\end{eqnarray}
The frequency $\omega$ represents the summation of all possible frequencies with different momentum modes i.e. $\omega= \sum_p \omega_p= \int \omega_p d^3p$. The factor $1/2$ as we deal with vacuum energy.  Substituting T= 40.96 K from Eq. (\ref{temperature}), the reduced collective wavelength is found to be
\begin{eqnarray}
    \lambdabar= 2.7 \times 10^{-5} m
\end{eqnarray}
This value aligns with the one we derived from the cutoff in Eq. (\ref{L_z}). This alignment serves as an affirmation of our theory that presents dark energy as a Bose-Einstein condensate, thereby fortifying our proposed resolution for the enigmatic cosmological constant's value. 
One might ask, what if we used a different distribution function, like a Gaussian distribution? Doing so would give a completely different temperature value and corresponding length scale. This tells us that the Bose-Einstein distribution has a very special role in figuring out the cutoff length scale found in relation to spacetime uncertainty. The spacetime uncertainty/cutoff in the energy density evaluations is now recognized as an inherent feature of the Bose-Einstein condensate which we ascribe to dark energy.  Present techniques that manipulate BECs with lasers and magnetic fields \cite{anderson1995observation} could furnish grounds for testable predictions about dark energy. The hypothesis of dark energy as a BEC, therefore, not only offers a stimulating perspective to understand this elusive component of our universe, but also delivers a novel avenue to investigate the fundamental properties of quantum fields on cosmological scales.

It is important to clarify that the massless bosons under consideration are not photons, but rather massless gluons. In a recent study \cite{Inan:2024noy}, we demonstrated that dark energy exhibits properties akin to those of a superconductor. It is noteworthy that superconductivity typically manifests in materials near zero Kelvin, yet we have observed such behavior at 41 K, as detailed in our study. Superconductivity is characterized by zero electrical resistance and the expulsion of magnetic fields, phenomena mathematically described by the breaking of U(1) symmetry, akin to the experimental Meissner effect. Recent studies, including one by one of  authors \cite{Ali:2024rnw}, have identified the residual symmetry in superconducting materials as SU(3). This is reflective of the fundamental forces structured as \(SU(3) \times SU(2) \times U(1)\) in nature. At the electroweak scale, \(SU(2)\) symmetry breaks, and near zero Kelvin, \(U(1)\) symmetry breaks, leaving \(SU(3)\) as the dominant unbroken symmetry. Thus, at near zero Kelvin, the vacuum is predominantly characterized by \(SU(3)\) symmetry. This suggests that gluons, being massless gauge bosons, could be candidates for dark energy and might exhibit Bose-Einstein condensation (BEC). However, as previously identified, gluons may not adhere strictly to BEC, necessitating a more precise statistical distribution to reconcile the gap between the BEC-derived temperature of 41 K and the observed temperature of the CMB at 2.7 K.

The Cosmic Microwave Background (CMB) temperature is different for various massless particles~\cite{Singleton:2015dqa}. For example, for photons it is approximately $2.7~\mathrm{K}$. For neutrinos, it is around $2.0~\mathrm{K}$. For gravitons, it is estimated to be about $1.0~\mathrm{K}$. However, none of these particle backgrounds can account for the observed cosmological expansion. We propose that, instead of invoking dark energy to explain this expansion, a Bose–Einstein condensate (BEC) can be considered. The CMB-like temperature associated with this condensate is determined by the energy density required to drive the observed rate of cosmic expansion. This temperature is approximately $41~\mathrm{K}$, and its experimental signature is precisely that it yields the observed Hubble expansion rate.

\section{Conclusion}

In this work, we have proposed a resolution to the cosmological constant problem by linking it to a fundamental uncertainty in the spacetime metric—a concept that emerges naturally from the interplay between general relativity and quantum mechanics. Experimental results have shown that gravitational effects remain operative down to length scales of order $10^{-5}$ m, and that quantum coherence and macroscopic quantum phenomena such as Bose-Einstein condensation also manifest at this scale. These observations motivate the extension of the principle that mass/energy warps spacetime to the quantum regime.

At the heart of our argument lies the realization that quantum uncertainty in particle localization leads to an intrinsic ambiguity in how spacetime is shaped at small scales. This metric uncertainty, quantified as $\Delta g \sim 10^{-61}$, translates into a natural cutoff scale $L_Z \sim 2.2 \times 10^{-5}$ m. Remarkably, this length scale is not arbitrary—it corresponds to the geometric mean between the Planck length and the cosmic horizon, suggesting a mesoscopic regime where quantum and gravitational effects coexist.

To support this interpretation, we employed two complementary approaches. The first derives $L_Z$ directly from the observed vacuum energy density, showing that it naturally emerges when spacetime uncertainty is accounted for in QFT. The second approach models dark energy as a Bose-Einstein condensate of massless bosons. Applying the Bose-Einstein distribution yields a thermal energy density with a characteristic wavelength that coincides with the same cutoff length $L_Z$, offering a thermodynamic interpretation of the vacuum.

The consistency between these two approaches underscores the physical significance of the cutoff scale. We find that other statistical distributions (e.g., Gaussian) do not reproduce this value, highlighting the privileged role played by Bose-Einstein statistics in linking quantum coherence with cosmic vacuum energy. Furthermore, our framework suggests that the Bose-Einstein condensate temperature (approximately 41 K) is not a prediction of the cosmic microwave background, but rather an effective property of the dark sector condensate.

In conclusion, the convergence of gravitational, quantum, and thermodynamic reasoning around a single mesoscopic length scale suggests a deeper principle at work—one in which the large-scale structure of the universe is intimately tied to the microscopic uncertainty of quantum geometry. This perspective opens new avenues for connecting macroscopic quantum states with cosmological observables, and may lay the foundation for a revised understanding of QFT in curved spacetime.

\section*{Compliance with Ethical Standards}
Authors have no conflict of interest to declare. The authors have no competing interests to declare that are relevant to the content of this article.

\section*{Data Availability Statement}

No Data associated in the manuscript.

\end{document}